\documentclass[twocolumn,showpacs,preprintnumbers,amsmath,amssymb,pra]{revtex4-1}

\usepackage{graphicx}
\usepackage{dcolumn}
\usepackage{bm}
\usepackage{color}

\begin{document}

\title{Universal quantum computation with a nonlinear oscillator network}

\author{Hayato Goto}
\affiliation{Frontier Research Laboratory, 
Corporate Research \& Development Center, 
Toshiba Corporation, 
1, Komukai Toshiba-cho, Saiwai-ku, Kawasaki-shi, 212-8582, Japan }

\begin{abstract}

It has recently been shown that
a parametrically driven oscillator with Kerr nonlinearity
yields a Schr\"odinger cat state via quantum adiabatic evolution through its bifurcation point
and a network of such nonlinear oscillators can be used for solving combinatorial optimization problems
by bifurcation-based adiabatic quantum computation
[H. Goto, Sci. Rep. \textbf{6}, 21686 (2016)].
Here we theoretically show that such a nonlinear oscillator network with controllable parameters
can also be used for universal quantum computation.
The initialization is achieved by a quantum-mechanical bifurcation
based on quantum adiabatic evolution,
which yields a Schr\"odinger cat state.
All the elementary quantum gates are also achieved by quantum adiabatic evolution,
in which dynamical phases accompanying the adiabatic evolutions are controlled 
by the system parameters.
Numerical simulation results indicate that 
high gate fidelities can be achieved,
where no dissipation is assumed.

\end{abstract}


\maketitle

\textit{Introduction}.
The standard model for quantum computation
consists of quantum bits (qubits) and quantum gates
\cite{Nielsen}
as present-day digital computers consist of bits and logic gates.
A qubit is often represented by two discrete quantum states 
of various physical systems, such as 
electron or nuclear spins in neutral atoms, ions, molecules, or solids,
polarization states or optical modes of single photons, 
and superconducting artificial atoms with Josephson junctions
\cite{Ladd2010a}.
Another kind of implementation of a qubit uses a harmonic oscillator,
which is described by an infinite-dimensional Hilbert space.
In this case,
the two computational basis states are defined as 
two orthogonal states of a harmonic oscillator,
such as two coherent states with largely different amplitudes 
or two cat states with opposite parity
\cite{note1}.
It is known that
a universal gate set for such coherent-state qubits
can be achieved by gate teleportation with cat states
\cite{Ralph2003a,Lund2008a}.
Recently,
universal quantum computation with harmonic oscillators
accompanied by nonlinear losses has also been proposed
\cite{Mirrahimi2014a,Albert2016a}.
Recent advances in circuit quantum electrodynamics 
with superconducting devices \cite{Vlastakis2013a,Leghtas2015a}
make the proposals promising.

More recently,
it has been shown that
a parametrically driven oscillator with Kerr nonlinearity 
(hereafter KPO) can yield a cat state 
via its quantum-mechanical bifurcation based on quantum adiabatic evolution,
and a network of such nonlinear oscillators
can be used for adiabatic quantum computation to find
the ground states of the Ising model
\cite{Goto2016a}.
In the bifurcation-based adiabatic quantum computation,
two coherent states, $|\alpha \rangle $ and $|-\alpha \rangle$, 
corresponding to two stable branches of the KPO
are regarded as up and down states of an Ising spin.
It may be natural to expect that the two coherent states of a KPO
can also be utilized as a qubit in the standard gate-based model of quantum computation.
However, it has not been obvious so far whether 
a universal gate set can be achieved for such qubits with KPOs.
(Note that no quantum gates are used in the bifurcation-based adiabatic quantum computation,
in which the necessary operation is only to increase pump amplitudes monotonically.
It should also be noted that the proposals in Refs. \cite{Mirrahimi2014a,Albert2016a}
cannot be directly applied to the KPOs
because of the differences in their nonlinearities.)

\begin{figure}[htbp]
\begin{center}
	\includegraphics[width=5.5cm]{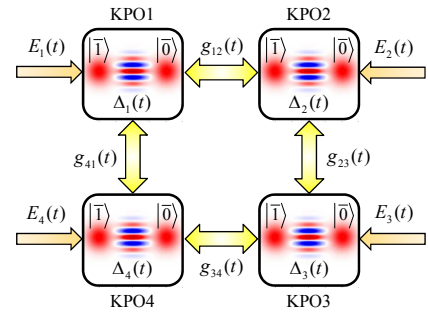}
	\caption{
	Schematic of the proposed quantum computer.
	The $R_z$ gate in Eq. (\ref{eq-Rz}) is performed by a driving field denoted by $E(t)$.
	The $R_x$ gate in Eq. (\ref{eq-Rx}) is performed by controlling a detuning denoted by $\Delta(t)$.
	The two-qubit gate in Eq. (\ref{eq-U}) is performed by controlling 
	a coupling coefficient denoted by $g(t)$.
	Each KPO is represented by the Wigner function of a cat state.
	The coherent states with positive and negative amplitudes
	correspond to the two computational basis states 
	$|\bar{0} \rangle$ and $|\bar{1} \rangle$, respectively.}
	\label{fig-schematic}
\end{center}
\end{figure}

In this Rapid Communication, 
we theoretically show that a universal gate set can be achieved for
the above qubits with KPOs.
Figure \ref{fig-schematic} shows a schematic of the proposed quantum computer.
All the elementary gates are based on quantum adiabatic evolution,
in which dynamical phases accompanying the adiabatic evolutions 
are controlled by system parameters.
In the following, we first describe
the definition of qubits in the present model.
The physical implementation of the KPO is also mentioned.
Next, we explain how to perform
three kinds of elementary gates on the qubits.
Numerical simulation results supporting the proposal
are also provided.
Finally, the conclusion is presented.
Note that in the present work, we assume that
there are no control errors and no decoherence sources such as losses.
Such errors will be considered in a future work.

\textit{Definition of qubits}.
In the present model,
we use a KPO for each qubit.
Here we first describe the cat-state generation with a KPO 
via quantum adiabatic evolution,
and next explain the definition of the qubit.

In a frame rotating at half the pump frequency of 
the parametric drive and in the rotating-wave approximation, 
the Hamiltonian for a KPO is given by
\begin{align}
H_1=\hbar \Delta a^{\dagger} a + \hbar \frac{K}{2} a^{\dagger 2} a^2
- \hbar \frac{p}{2} (a^{\dagger 2} + a^2),
\label{eq-Hamiltonian1}
\end{align}
where 
$a$ and $a^{\dagger}$ are the annihilation and creation operators
for the KPO,
$\Delta$ is the detuning of the eigenfrequency from half the pump frequency, 
$K$ is the Kerr coefficient for the Kerr effect,
and $p$ is the pump amplitude for the parametric drive \cite{Goto2016a}.
Hereafter, 
we assume that $K$ is a positive constant and $\Delta$ is nonnegative.
(When $K$ is negative, similar discussion is straightforward.)

A cat state of the KPO is generated deterministically as follows.
The KPO is initially prepared in the vacuum state $|0\rangle$.
Then, $p$ is increased sufficiently slowly from zero.
Since $|0\rangle$ is the ground state for the initial Hamiltonian,
the KPO adiabatically follows the instantaneous ground state of the Hamiltonian.
When $\Delta =0$,
the ground state is doubly degenerate 
and the eigenspace is spanned by two coherent states $|\pm \sqrt{p/K} \rangle$.
Since $H_1$ is symmetric under parity inversion $a\to -a$,
the final state should have the same parity as the initial state $|0\rangle$.
Thus, the final state is the even cat state \cite{note1} defined as
\begin{align}
|C_+ \rangle = \frac{|\sqrt{p/K} \rangle + |-\sqrt{p/K} \rangle }{\sqrt{2(1+e^{-2p/K})}}.
\label{eq-cat}
\end{align}

When $p$ is so large compared to $K$ that $e^{-2p/K}$ is negligible,
the two coherent states $|\pm \sqrt{p/K} \rangle$ are orthogonal
to each other.
Taking a value $p_0$ such that $e^{-2p_0/K}$ is negligible,
we define the computational basis states, $|\bar{0} \rangle$ and $|\bar{1} \rangle$, 
of a qubit 
as two coherent states $|\sqrt{p_0/K} \rangle$ and  $|- \sqrt{p_0/K} \rangle$,
respectively,
where we have used the bars to distinguish the computational basis states 
from the vacuum and single-photon states of the KPO.
(In this Rapid Communication, the quanta of the KPO are called ``photons"
assuming that the KPO is implemented by an electromagnetic resonator.)

Note that the cat-state generation described above is regarded as
the initialization of the qubit to 
$(|\bar{0} \rangle + |\bar{1} \rangle )/\sqrt{2}$,
which is the standard initial state in quantum computation
\cite{Nielsen,Grover1996a,Shor1997a}.
After this initialization,
the pump amplitude is kept to $p_0$ during quantum computation.
While no operation is performed,
the state of each KPO is 
in the subspace spanned by $|\bar{0} \rangle$ and $|\bar{1} \rangle$.

Here we briefly address the physical implementation of the KPO.
Since the time scale is limited by the Kerr coefficient $K$,
a large $K$ is desirable.
(As shown below, the gate time is proportional to $K^{-1}$.)
In particular, it is desirable that
$K$ should be larger than the loss rate of the KPO.
This condition is extremely stringent for optical systems.
On the other hand, superconducting systems with Josephson junctions
have already achieved this condition
\cite{Kirchmair2013a,Rehak2014a}.
With a superconducting circuit,
parametric oscillation has also been demonstrated 
\cite{Lin2014a}.
Thus, superconducting systems are promising for the implementation of the KPO.
In this case, 
$K$ is typically several tens of MHz.

\textit{Elementary quantum gates}.
Here we show that a universal gate set can be achieved 
for the qubit defined above.
As a universal gate set, 
we choose two single-qubit gates
$R_z(\phi)$ and $R_x (\theta) $, and
a two-qubit gate $U(\Theta )$,
where these unitary operators are defined as follows
($X$ and $Z$ denote Pauli operators)
\cite{Nielsen}:
\begin{align}
R_z(\phi)&
=e^{-i\phi Z/2}
=
\begin{pmatrix}
e^{-i\phi/2} & 0
\\
0 & e^{i\phi/2} 
\end{pmatrix},
\label{eq-Rz}
\\
R_x(\theta)
&=e^{-i\theta X/2}
=
\begin{pmatrix}
\cos \frac{\theta}{2} & -i\sin \frac{\theta}{2}
\\
-i\sin \frac{\theta}{2} & \cos \frac{\theta}{2} 
\end{pmatrix},
\label{eq-Rx}
\\
U(\Theta)&
=e^{-i\Theta Z_1 Z_2 /2}
=
\begin{pmatrix}
e^{-i\Theta/2} & 0 & 0 & 0
\\
0 & e^{i\Theta/2} & 0 & 0 
\\
0 & 0 & e^{i\Theta/2} & 0
\\
0 & 0 & 0 & e^{-i\Theta/2}
\end{pmatrix}.
\label{eq-U}
\end{align}
In the following, we explain how to perform these elementary gates in turn.
[Note that $R_x(\pi /2)$ and $U(\pi /2)$ are sufficient for universality
together with $R_z(\phi)$ \cite{note2}.]

To perform $R_z(\phi)$ on a qubit,
we drive the KPO by a driving field with a pulse-shaped amplitude
$E(t)$.
Then, the additional Hamiltonian is given by
\begin{align}
H_{z}(t) = \hbar E(t) (a+a^{\dagger}).
\label{eq-Ein-Hamiltonian}
\end{align}
Note that the additional Hamiltonian violates
the parity symmetry, and consequently induces
the transition between the even and odd cat states.
This driving is also interpreted as a displacement in the phase space.

When $|E(t)|$ is sufficiently small and the variation of $E(t)$ is sufficiently slow,
the KPO is approximately kept in the subspace spanned by $|\bar{0} \rangle$ and $|\bar{1} \rangle$.
Here we should consider the energy shifts for $|\bar{0} \rangle$ and $|\bar{1} \rangle$,
which are $2\hbar E(t) \sqrt{p_0/K}$ and 
$-2\hbar E(t) \sqrt{p_0/K}$, respectively.
These energy shifts induce dynamical phase factors, and consequently
$R_z(\phi)$ is performed,
where $\phi$ is given by ($T_g$ is the gate time)
\begin{align}
\phi = 4 \sqrt{p_0/K} \int_0^{T_g} E(t) dt.
\label{eq-phi}
\end{align}

To verify the above discussion, we did numerical simulations,
in which we numerically solved the Schr\"odinger equation
with the Hamiltonian $H_1+H_z (t)$.
In the simulations, the parameters are set as $p_0=4K$, $\Delta =0$, and $T_g=2/K$,
and the Hilbert space is truncated at a photon number of 20.
We performed $R_z(\phi )$ on the initial state $(|\bar{0} \rangle + |\bar{1} \rangle )/\sqrt{2}$
and calculated the fidelity between the output state in the simulation and 
the ideal output state $(e^{-i\phi/2}|\bar{0} \rangle + e^{i\phi/2}|\bar{1} \rangle )/\sqrt{2}$.
(The fidelity is defined as the square of the absolute value of the inner product of two state vectors.)
To perform $R_z (\phi )$,
$E(t)$ is set as
\begin{align}
E(t)=\frac{\pi \phi}{8T_g\sqrt{p_0/K}} \sin \frac{\pi t}{T_g}.
\label{eq-Ein}
\end{align}

\begin{figure}[htbp]
\begin{center}
	\includegraphics[width=7cm]{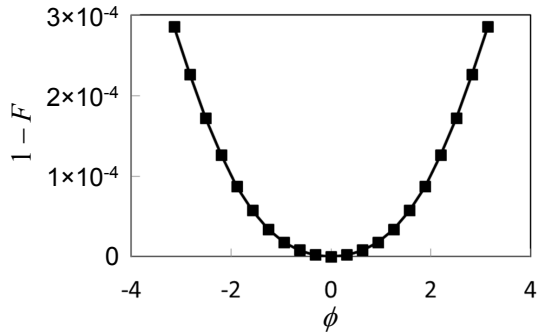}
	\caption{
	Simulation result for $R_z(\phi)$.
	$F$ denotes the fidelity between the output state in the simulation and 
	the ideal output state $(e^{-i\phi/2}|\bar{0} \rangle + e^{i\phi/2}|\bar{1} \rangle )/\sqrt{2}$.}
	\label{fig-Rz}
\end{center}
\end{figure}

The simulation result is shown in Fig. \ref{fig-Rz}.
It is found that 
high fidelities are achieved for $\phi$ in the range of $-\pi$ to $\pi$, as expected.
The fidelities become even higher for a longer gate time
(a larger value of $KT_g$).

Next, we explain how to perform $R_x(\theta)$.
To perform $R_x(\theta)$ on a qubit,
we use the detuning $\Delta $ in Eq. (\ref{eq-Hamiltonian1}). 
When $\Delta$ is slowly increased from zero to a value $\Delta_0$ near to $p_0$
and then decreased to zero,
the even and odd cat states,
$(|\bar{0} \rangle \pm |\bar{1} \rangle )/\sqrt{2}$,
obtain dynamical phase factors depending on their energy shifts
due to the nonzero detuning.
(Note that the even and odd states are simultaneous eigenstates of $H_1$ 
because of the parity symmetry of  $H_1$.)
Thus, the qubit state changes as follows
($\theta$ is the relative phase
between the even and odd cat states
due to the dynamical phase factors):
\begin{widetext}
\begin{align}
\alpha_0 |\bar{0}\rangle + \alpha_1 |\bar{1}\rangle
&=
\frac{\alpha_0 + \alpha_1}{2} (|\bar{0}\rangle +|\bar{1}\rangle)
+
\frac{\alpha_0 - \alpha_1}{2} (|\bar{0}\rangle -|\bar{1}\rangle)
\nonumber \\
&\to
\frac{\alpha_0 + \alpha_1}{2} (|\bar{0}\rangle +|\bar{1}\rangle)
+
\frac{\alpha_0 - \alpha_1}{2} e^{i\theta} (|\bar{0}\rangle -|\bar{1}\rangle)
\nonumber \\
&=
e^{-i\theta/2}
\left[
\left(
\alpha_0 \cos \frac{\theta}{2} -i\alpha_1 \sin \frac{\theta}{2}
\right) |\bar{0}\rangle
+
\left(
\alpha_1 \cos \frac{\theta}{2} -i\alpha_0 \sin \frac{\theta}{2}
\right) |\bar{1}\rangle
\right]
\nonumber \\
&= e^{-i\theta/2}
R_x(\theta) (\alpha_0 |\bar{0}\rangle + \alpha_1 |\bar{1}\rangle).
\label{eq-theta}
\end{align}
\end{widetext}
Thus, $R_x(\theta)$ is achieved by the detuning control.
(The overall phase factor $e^{-i\theta/2}$ has no physical meaning
and therefore can be ignored.)

Here we present numerical simulation results supporting the above discussion.
We numerically solved the Schr\"odinger equation
with $H_1$ in Eq. (\ref{eq-Hamiltonian1}),
where the detuning $\Delta$ is controlled as follows
($T_g$ is the gate time):
\begin{align}
\Delta (t) = \Delta_0 \sin^2 \frac{\pi t}{T_g}.
\label{eq-Delta}
\end{align}
In the simulations, the parameters are set as $p_0=4K$ and $T_g=10/K$,
the initial state $|\psi_i \rangle$ is set to $(|\bar{0} \rangle + i|\bar{1} \rangle )/\sqrt{2}$,
and the Hilbert space is truncated at a photon number of 20.

\begin{figure}[htbp]
\begin{center}
	\includegraphics[width=7cm]{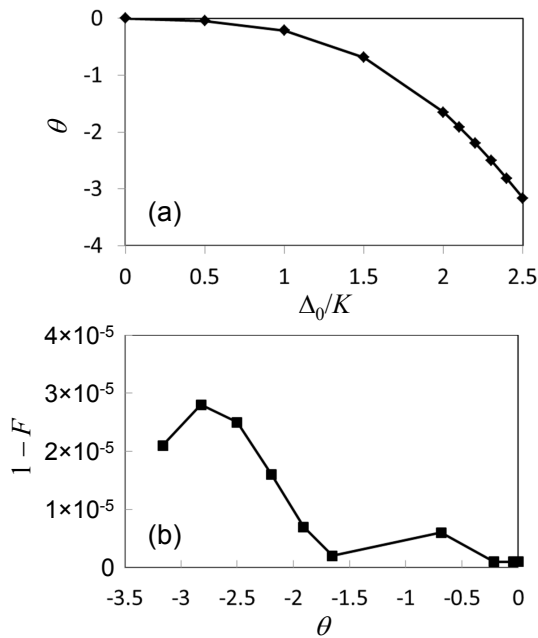}
	\caption{
	Simulation result for $R_x(\theta)$.
	(a) $\theta$ maximizing the fidelity between the output state in the simulation
	and $R_x(\theta) |\psi_i \rangle$.
	(b) $F$ denotes the fidelity between the output state in the simulation 
	and $R_x(\theta) |\psi_i \rangle$.}
	\label{fig-Rx}
\end{center}
\end{figure}

To estimate the rotation angle $\theta$ corresponding to $\Delta_0$,
we calculated the fidelity between the output state in the simulation
and $R_x(\theta) |\psi_i \rangle$, and found $\theta$ maximizing the fidelity.
The results are summarized in Fig. \ref{fig-Rx}, where
$F$ denotes the maximized fidelity.
While $\Delta_0$ changes from 0 to $2.5K$,
the rotation angle $\theta$ changes from 0 to $-\pi$.
The fidelities are also high.
Thus, it has been shown that $R_x(\theta)$ is achieved by controlling the detuning.

\begin{figure}[htbp]
\begin{center}
	\includegraphics[width=7cm]{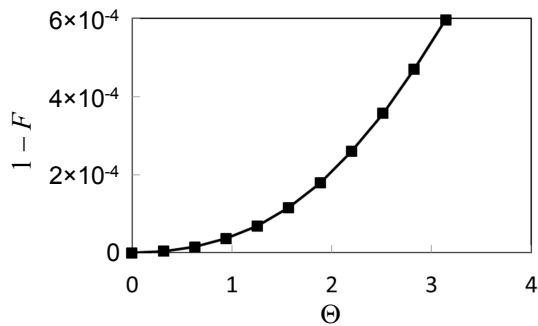}
	\caption{
	Simulation result for the two-qubit gate $U(\Theta)$.
	$F$ denotes the fidelity between the output state in the simulation and 
	the ideal output state 
	$U(\Theta)|\psi_i \rangle = 
	(e^{-i\Theta/2}|\bar{0} \rangle |\bar{0} \rangle 
	+ e^{i\Theta/2} |\bar{0} \rangle |\bar{1} \rangle 
	+ e^{i\Theta/2} |\bar{1} \rangle |\bar{0} \rangle 
	+ e^{-i\Theta/2} |\bar{1} \rangle |\bar{1} \rangle )/2$.}
	\label{fig-U}
\end{center}
\end{figure}

Finally, we explain how to perform the two-qubit gate $U(\Theta)$.
To perform $U(\Theta)$ on two qubits,
we control the coupling coefficient $g(t)$ between the two KPOs.
(In contrast, in the bifurcation-based adiabatic quantum computation for the Ising problem
\cite{Goto2016a},
the coupling coefficients are set to constants depending on given problems.)
The additional Hamiltonian is given by
\begin{align}
H_U=\hbar g(t) (a_1 a_2^{\dagger}+a_1^{\dagger} a_2).
\label{eq-g-Hamiltonian}
\end{align}
Note that this is the standard \textit{linear} coupling,
which describes photon exchange between two KPOs.

When $|g(t)|$ is sufficiently small and the variation of $g(t)$ is sufficiently slow,
the KPOs are approximately kept in the subspace spanned by 
$|\bar{0} \rangle |\bar{0} \rangle$, 
$|\bar{0} \rangle |\bar{1} \rangle$,
$|\bar{1} \rangle |\bar{0} \rangle$, and $|\bar{1} \rangle |\bar{1} \rangle$.
Then, the energy shifts for $|\bar{0} \rangle |\bar{0} \rangle$ and 
$|\bar{1} \rangle |\bar{1} \rangle$
are $2\hbar g(t) p_0/K$ and 
those for $|\bar{0} \rangle |\bar{1} \rangle$ and 
$|\bar{1} \rangle |\bar{0} \rangle$
are $-2\hbar g(t) p_0/K$.
These energy shifts induce dynamical phase factors, and consequently
$U(\Theta)$ is performed,
where $\Theta$ is given by ($T_g$ is the gate time)
\begin{align}
\Theta = \frac{4p_0}{K} \int_0^{T_g} g(t) dt.
\label{eq-Theta}
\end{align}

We did numerical simulations for the two-qubit gate,
in which we numerically solved the Schr\"odinger equation.
In the simulations,
the parameters are set as $p_0=4K$, $\Delta =0$, and $T_g=2/K$,
the initial state is set to
$|\psi_i \rangle =
(|\bar{0} \rangle + |\bar{1} \rangle )(|\bar{0} \rangle + |\bar{1} \rangle )/2$,
and the Hilbert space is truncated at a photon number of 20 for each KPO.
We performed $U(\Theta )$ on $|\psi_i \rangle$ 
and calculated the fidelity between the output state in the simulation and 
the ideal output state $U(\Theta)|\psi_i \rangle$.
To perform $U (\Theta )$,
$g(t)$ is set as
\begin{align}
g(t)=\frac{\pi \Theta}{8T_g p_0/K} \sin \frac{\pi t}{T_g}.
\label{eq-g}
\end{align}

The simulation result is shown in Fig. \ref{fig-U}.
It is found that 
high fidelities are achieved for $\Theta$ in the range of 0 to $\pi$, as expected.
The fidelities become even higher for a longer gate time
(a larger value of $KT_g$).

In conclusion,
we have shown that
the network of nonlinear oscillators called KPOs,
which has been used for bifurcation-based adiabatic quantum computation,
can also be used for universal quantum computation.
The qubit is defined with two coherent states of a KPO.
The initialization of the qubit is achieved by 
a quantum-mechanical bifurcation based on quantum adiabatic evolution.
All the elementary gates are also performed by quantum adiabatic evolution,
in which dynamical phases accompanying the adiabatic evolutions
are controlled by driving field amplitudes, detunings, and coupling coefficients.
The present scheme will open a new possibility for 
quantum nonlinear systems in quantum information science.


\begin{thebibliography}{19}
\bibitem{note1}
A coherent state $|\alpha \rangle$ is defined 
as an eigenstate of the annihilation (lowering) operator, $a$,
of the harmonic oscillator:
$a|\alpha \rangle = \alpha|\alpha \rangle$.
Since the inner product of two coherent states 
$|\alpha \rangle$ and $|\beta \rangle$ is given by
$\langle \alpha | \beta \rangle = \exp[-|\alpha - \beta |^2 /2 +i\mbox{Im}(\alpha^* \beta)]$,
two coherent states with largely different amplitudes 
are approximately orthogonal.
And even and odd cat states are defined as 
$|\alpha \rangle + |-\alpha \rangle$ and
$|\alpha \rangle - |-\alpha \rangle$, respectively,
where the normalization factors have been omitted.
\bibitem{note2}
$e^{i\pi /2} R_z(\pi /2) R_x(\pi /2) R_z(\pi /2)$ is equivalent to a Hadamard gate.
$e^{-i\pi /4} U(\pi /2) R_z^{(1)}(-\pi /2) R_z^{(2)}(-\pi /2) $
is equivalent to a controlled-$Z$ gate,
where $R_z^{(1)}(\phi) = e^{-i\phi Z_1/2}$ and $R_z^{(2)}(\phi) = e^{-i\phi Z_2/2}$.
\end{thebibliography}
\end{document}